\documentclass[12pt,a4paper]{iopart}
\usepackage{iopams}  

\usepackage{graphicx}
\usepackage{amssymb} 
\usepackage{float} 
\usepackage{subfig} 
\usepackage{times}
\usepackage{txfonts}

\usepackage{ifpdf}
\ifpdf
\usepackage[pdftex]{hyperref}
\else
\usepackage[hypertex]{hyperref}
\fi

\hypersetup{
  pdftitle={},%
  pdfauthor={},%
  pdfsubject={},%
  pdfkeywords={},%
  pdfstartview={},%
  bookmarksopen=true, breaklinks=true, debug=true, %
  colorlinks=true, linkcolor=red, citecolor=blue, urlcolor=blue
}

\newcommand{\beq}{\begin{eqnarray}}
\newcommand{\eeq}{\end{eqnarray}}
\newcommand{\be}{\begin{eqnarray*}}
\newcommand{\ee}{\end{eqnarray*}}

\newcommand{\ie}{{\it i.e.}}

\newcommand{\Br}{{\rm Br}}

\newcommand{\cf}[1]{{Fig.~\ref{#1}}}

\def\lsim{\raise0.3ex\hbox{$<$\kern-0.75em\raise-1.1ex\hbox{$\sim$}}}
\def\gsim{\raise0.3ex\hbox{$>$\kern-0.75em\raise-1.1ex\hbox{$\sim$}}}

\def\pp   {$pp$}

\def\jpsi   {\mbox{$J/\psi$}}

\def\beq     {\begin{equation}}
\def\eeq     {\end{equation}}

\long\def\symbolfootnote[#1]#2{\begingroup%
  \def\thefootnote{\fnsymbol{footnote}}\footnote[#1]{#2}\endgroup}

\graphicspath{{plots/}}

\begin{document}

%

\title[$J/\psi$ production at $\sqrt{s}=1.96$ and  $7$ TeV]{$J/\psi$ production at $\sqrt{s}=1.96$ and  $7$ TeV: Color-Singlet Model, NNLO$^\star$ and polarisation}

\author[J.~P.~Lansberg]{J.~P.~Lansberg}

\address{IPNO, Universit\'e Paris-Sud 11, CNRS/IN2P3, 91406 Orsay, France}

\ead{Jean-Philippe.Lansberg@in2p3.fr}

\begin{abstract}
We study $J/\psi$ production   in $pp$ collisions at $\sqrt{s}=1.96$ and  $7$ TeV
 using the Colour-Singlet Model (CSM), 
including next-to-leading order (NLO) corrections and  dominant $\alpha_S^5$ 
contributions (NNLO$^\star$).  We find that the CSM reproduces the existing data if the upper
range of the NNLO$^\star$ is near the actual --but presently unknown -- NNLO. 
The direct yield polarisation for the NLO and NNLO$^\star$ 
is increasingly longitudinal in the helicity frame  when 
$P_T$ gets larger.  Contrary to what is sometimes claimed in the literature, 
the prompt $J/\psi$ yield polarisation in the CSM is compatible with 
the experimental data from the CDF collaboration,
when one combines the direct yield with a data-driven range for the polarisation
of $J/\psi$ from $\chi_c$.
\end{abstract}



\vspace*{-0.5cm}

\section{Introduction}
\label{sec:intro}

The numerous quarkonium-production puzzles at hadron colliders were attributed not too long ago to 
non-perturbative effects associated with channels in which the heavy
quark-pair is produced in a colour-octet state~\cite{Lansberg:2006dh}.
$\alpha^4_S$ and $\alpha^5_S$ corrections to the CSM~\cite{CSM_hadron} 
are now widely recognised as essential  to understand
the $P_T$ spectrum of $J/\psi$ and $\Upsilon$ produced in
high-energy hadron  collisions~\cite{Campbell:2007ws,Artoisenet:2007xi,Gong:2008sn,Artoisenet:2008fc,Lansberg:2008gk}.
This calls for a factorised description of high-$P_T$ $J/\psi$ beyond leading power~\cite{Kang:2011zz}. 
The effect of QCD corrections is also manifest in the polarisation predictions. While the 
$J/\psi$ and $\Upsilon$  produced inclusively or 
in association with a photon are predicted to be 
transversely polarised at LO, it has been found that their polarisation at NLO is 
 increasingly longitudinal when $P_T$ gets larger \cite{Gong:2008sn,Artoisenet:2008fc,Li:2008ym}.
In recent works~\cite{Brodsky:2009cf,Lansberg:2010cn}, we have also shown that the CSM  alone is 
sufficient to account for the  magnitude of $d\sigma/dy$ at RHIC, Tevatron and LHC energies.

We evaluate here the $P_T$ dependence 
of the \jpsi\ yield and its polarisation at Tevatron and LHC energies. 
We describe the procedure used to obtain a first evaluation of some dominant contributions at $\alpha_S^5$ (NNLO$^\star$)
in addition to the  yield at NLO (up to $\alpha_S^4$). We then compare available data from the Tevatron and the LHC with our results: the direct yields differential in  $P_T$ along with the polarisation vs $P_T$ for
the prompt yield using an essentially data-driven estimation of the polarisation for $J/\psi$ from $\chi_c$.

\section{Cross-section}

For the NLO cross section, we use the partonic matrix elements of~\cite{Campbell:2007ws}. 
In order to investigate the expected impact of NNLO QCD corrections for increasing
$P_T$, we also present the NLO results plus
the real-emission contributions at $\alpha_S^5$ evaluated along the lines of~\cite{Artoisenet:2008fc}, referred to as
NNLO$^\star$.  At $\alpha_S^5$, the 
last\footnote{We do not expect any further kinematical enhancement as regards the $P_T$ dependence when going
 further in the $\alpha_S$ expansion: $P_T^{-4}$ is the slowest possible fall-off. Above $\alpha_S^5$, usual 
expectations for the impact of QCD corrections would then hold. One would expect a $K$ factor multiplying 
the yield at NNLO accuracy, which would be independent of $P_T$ and of a similar size as those of other QCD processes.
A further enhancement by an order of magnitude between the NNLO and N$^3$LO results
would be quite worrisome.}   kinematically-enhanced topologies open up, with
a $P_T^{-4}$ fall off of $d\sigma/dP_T^2$. 
The procedure used here for the NNLO$^\star$  is exactly that of~\cite{Artoisenet:2008fc}: the 
real-emission contributions at $\alpha_S^5$ are evaluated using {\small MADONIA}~\cite{Artoisenet:2007qm} by imposing a lower bound on the invariant-mass squared of 
any light partons ($s_{ij}$). 
The dependence on this cut should decrease for larger $P_T$ since no collinear or soft divergences can appear there 
for the new channels opening up at $\alpha^5_S$ with a leading-$P_T$ behaviour, \ie~the ones which interest us.
For other channels, whose Born contribution is at $\alpha^3_S$ or $\alpha^4_S$, the cut would 
produce logarithms of $s_{ij}/s_{ij}^{\rm min}$. These are not necessarily small, but
they are expected to be factorised over their corresponding Born contribution, 
which scales as $P_T^{-8}$ or $P_T^{-6}$. They are thus
suppressed by at least two powers of $P_T$ with respect of the leading-$P_T$ contributions ($P_T^{-4}$). The sensitivity on 
$s_{ij}^{\rm min}$ is  expected to be small at large $P_T$.

\begin{figure}[h!]
\begin{center}
{\includegraphics[width=0.34\columnwidth]{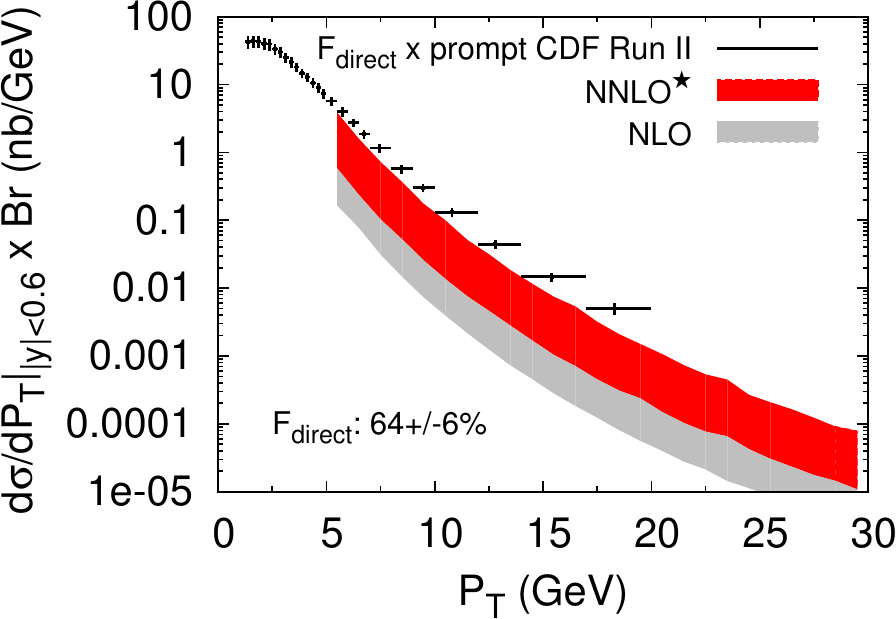}}
{\includegraphics[width=0.32\columnwidth]{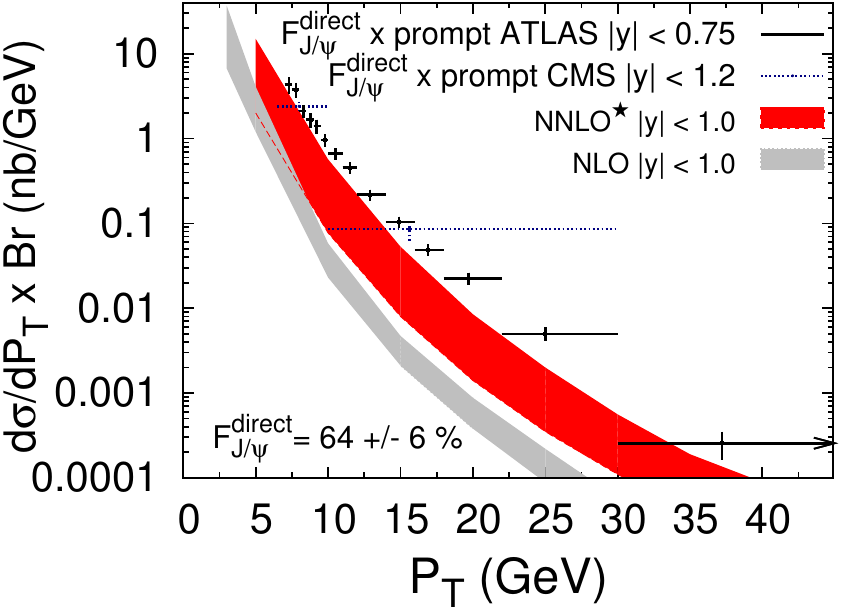}}
{\includegraphics[width=0.32\columnwidth]{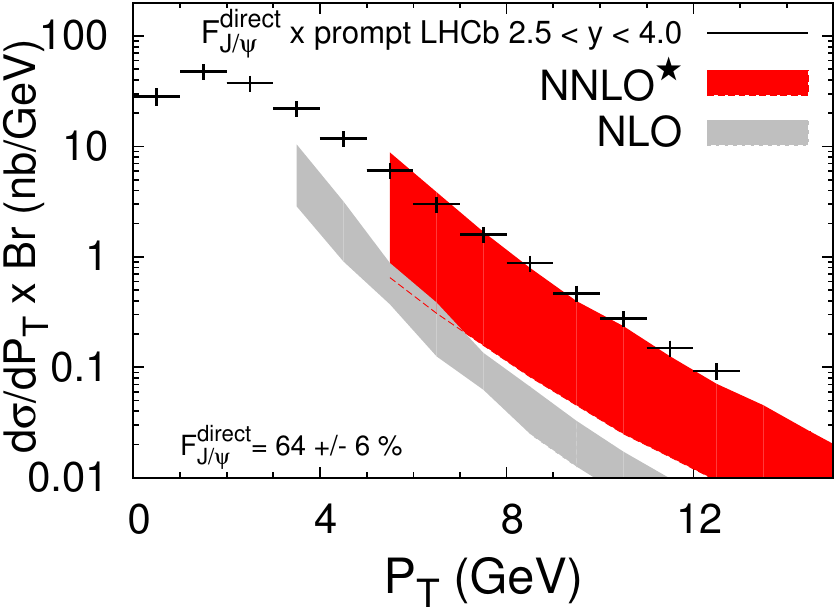}}
\end{center}
\caption{$d\sigma/dP_T \times {\rm Br} $ for direct $J/\psi$ production
from NLO and NNLO$^\star$ CS contributions  at $\sqrt{s}=1.96\mathrm{~TeV}$ (left)
and at $\sqrt{s}=7\mathrm{~TeV}$ for central  (middle) and forward (right) rapidities. These are
compared to the CDF~\cite{Acosta:2004yw}, ATLAS, CMS and LHCb data~\cite{Khachatryan:2010yr,ATLAS:2011sp,Aaij:2011jh}
multiplied by a constant direct fraction from CDF~\cite{Abe:1997yz}. See text for details on theoretical-error bands. }
\label{fig:dsigdpt}
\end{figure}

Our results are shown on \cf{fig:dsigdpt}. The CSM is very close to the existing data, if the upper
range of the NNLO$^\star$ is a relevant evaluation of the NNLO.  
The uncertainty bands at NLO are obtained from the {\it combined}
 variations of the charm-quark
mass ($m_c=1.5\pm 0.1$ GeV), the
factorisation $\mu_F$ and the renormalisation $\mu_R$ scales
 chosen in the couples $((0.75,0.75);(1,1);(1,2);(2,1);(2,2))\times m_T$ with $m^2_T=4m_Q^2+P_T^2$.
The band for the NNLO$^\star$ is obtained using a combined variation of $m_c$, $0.5 m_T <\mu_R=\mu_F< 2 m_T$
and $2.25  < s_{ij}^{\rm min}<   9.00$ GeV$^2$. We have used 
the NLO set {\small CTEQ6\_M}~\cite{Pumplin:2002vw} and have taken $|R_{J/\psi}(0)|^2=1.01$ GeV$^3$ and Br$(J/\psi \to \ell^+\ell^-)=0.0594$.  

\section{Polarisation}

The  polarisation parameter $\alpha$ is extracted bin by bin in $y$ or $P_T$ 
from the normalised 
distribution of the polar  angle $\theta$ between the $\ell^+$ direction in the
$J/\psi$ rest frame and its direction in the laboratory
frame, $I(\cos \theta) =
\frac{3}{2(\alpha+3)} (1+\alpha \, \cos^2 \theta)$. We thus work in the helicity frame.
$\alpha$ is also related to a ratio of the polarised
cross sections:
$\alpha=\frac{\sigma_T-2\sigma_L}{\sigma_T+2\sigma_L}$.

For the time being, there does not exist any measurement of direct $J/\psi$ polarisation, 
\ie~after the extraction of the $\chi_c$ feed-downs (up to 30-40  \%)  which
may strongly impact  on the observed values of $\alpha$. It is however possible to constrain 
its effects by using existing data on $\sigma_{\chi_{c1}}/\sigma_{\chi_{c2}}$ and by 
relying  on  E1  dominance for the transition $\chi_c\to J/\psi+\gamma$.

Indeed, using E1 dominance~\cite{Cho:1994gb}, one can obtain~\cite{Lansberg:2010vq} a  range of the
 yield of longitudinally (transversely) polarised $J/\psi$  in terms of
 simple relations involving the polarised $\chi_c$ yields.
Allowing for extreme cases, these relations allow the yield from $\chi_c$ to be fully transversely polarised, while
there is a minimal value of $\alpha$. Following the discussion of~\cite{Lansberg:2010vq} and taking
$R_{12}=\frac{\sigma_{\chi_{c1}} \Br(\chi_{c1}\to J/\psi \gamma)}
{ \sigma_{\chi_{c2}} \Br(\chi_{c2}\to J/\psi \gamma)}=2.5\pm 0.1$\cite{Abulencia:2007bra}, one obtains $\alpha^{min}_{\rm from~\chi{_c}}\simeq-0.42$, rather different than -1.

\begin{figure}[hbt!]
\begin{center}
\includegraphics[width=0.5\columnwidth]{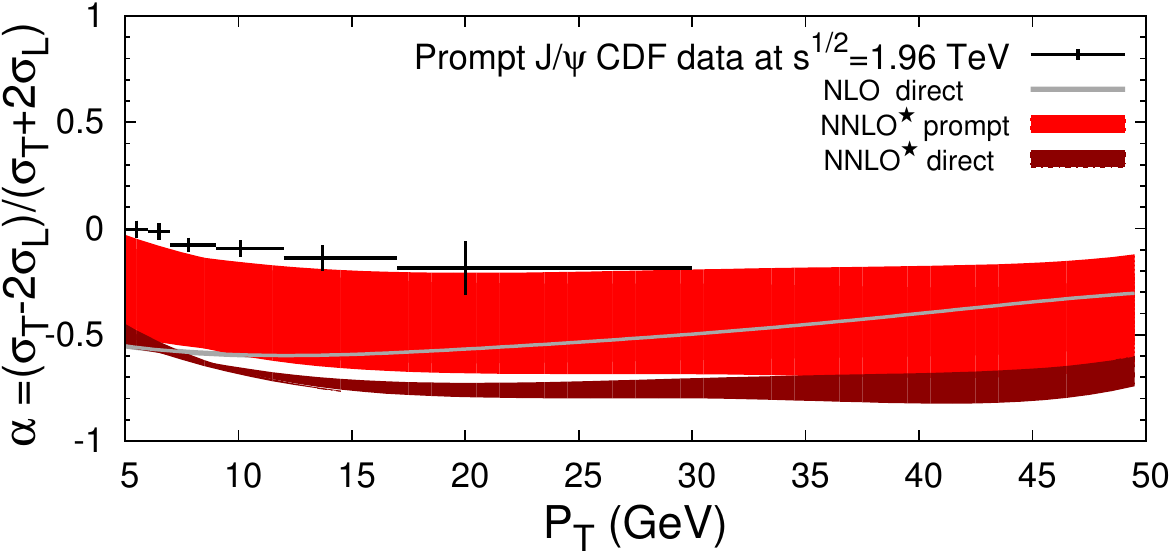}
\end{center}
\caption{Comparison between the extrapolation of $\alpha$ for prompt $J/\psi$ in \pp\ at $\sqrt{s}=1.96\mathrm{~TeV}$ (red band),   the direct NLO $\alpha$ (gray line), the direct NNLO$^\star$  $\alpha$  (thinner dark-red band) and the CDF 
data~\cite{Abulencia:2007us} for prompt $J/\psi$.}
\label{fig:alpha}
\end{figure}

Since  $30\%$ of the $J/\psi$ come from $\chi_c$ nearly independent of $P_T$ 
in the range considered here~\cite{Abe:1997yz}, we expect a partial contribution to the 
polarisation ranging from $0.3 \times (+1) $ to $0.3 \times(-0.42)$. Regarding the
other $70\%$, one multiplies the result for the direct yields 
by $0.7$, since the polarisation of $J/\psi$ from $\psi(2S)$ is expected to be 
identical to the direct one. Doing so, one obtains the extrapolation shown on~\cf{fig:alpha}. If the $J/\psi$
from $\chi_c$ yield is strongly transversely polarised (the upper limit), the polarisation of the prompt yield is in rather
good agreement with the data.

\section{Conclusion}

We have evaluated the NLO and  NNLO$^\star$ $J/\psi$ yield at
 Tevatron and LHC energies. As  found for $\Upsilon$ at the Tevatron~\cite{Artoisenet:2008fc} and for
 $J/\psi$ at RHIC~\cite{Lansberg:2010vq}, the upper bound of the CSM predictions is very close to the experimental
data from CDF, ATLAS, CMS and LHCb.  However, the NNLO$^\star$ evaluation is not a complete NNLO calculation. 
It is affected by  logs of an IR cut-off whose effect might not vanish as quickly as one has anticipated. It may very well be that 
the upper limit of the prediction --close to the data-- accurately reproduces  the complete NNLO yield, or that the lower limit
--close to the NLO yield-- reproduces the NNLO yield. 
If the upper limit of the NNLO$^\star$ does indeed overestimate the NNLO, the CSM alone is likely insufficient to 
account for the data. Conversely, 
 the CSM alone is enough and the colour-octet contributions are not required. 

As regards polarisation, we have derived a range for the prompt yield polarisation. This range is affected
by admittedly large theoretical uncertainties, but the upper edge -- corresponding to a transversely polarised feed-down-- is 
in rather good agreement with the data from the Tevatron. We recall that the trend for a longitudinally (direct) $\psi(2S)$ yield~\cite{Abulencia:2007us} was also met by the NNLO$^\star$~\cite{Lansberg:2008gk}.

In conclusion, the CSM may very well provide a good description of $J/\psi$ production in high-energy $pp$ collisions,
both in terms of the cross section and the yield polarisation.


\subsubsection*{Acknowledgments}

We thank P.~Artoisenet, J.~Campbell, F.~Maltoni and F.~Tramontano for our fruitful collaboration  
on~\cite{Artoisenet:2008fc}, from which the study presented here is derived.
We acknowledge the support of the ReteQuarkonii Networking of the EU I3 HP 2 program.

\end{document}